\documentclass{llncs}
\usepackage[style=base]{caption} 
\usepackage[misc,geometry]{ifsym}

\usepackage{graphics}
\usepackage{epsfig}
\usepackage{url}
\usepackage{hyperref}
\usepackage{hhline}
\usepackage[font=footnotesize,labelfont=bf]{caption}
\usepackage[shortlabels]{enumitem}

\usepackage{amsmath}

\usepackage{siunitx}
\usepackage{color}
\newcommand{\thomas}[1]{}
\renewcommand{\thomas}[1]{{\color{red}From Thomas: {#1}}}
\newcommand{\chris}[1]{}
\renewcommand{\chris}[1]{{\color{blue}From Chris: {#1}}}
\newcommand{\alan}[1]{}
\renewcommand{\alan}[1]{{\color{cyan}From Alan: {#1}}}

\usepackage{flushend}

\usepackage{algorithmic}
\usepackage[linesnumbered,ruled]{algorithm2e}

\usepackage{color}
\usepackage{xcolor}
\usepackage{listings,chngcntr}
\usepackage[framemethod=tikz]{mdframed}

\definecolor{mGreen}{rgb}{0,0.6,0}
\definecolor{mRed}{rgb}{0.6,0,0}
\definecolor{mBlue}{rgb}{0,0,0.6}
\definecolor{mGray}{rgb}{0.5,0.5,0.5}
\definecolor{mPurple}{rgb}{0.58,0,0.82}
\definecolor{backgroundColour}{rgb}{0.92,0.92,0.92}
\definecolor{light-gray}{gray}{0.95}
\definecolor{ballblue}{rgb}{0.13, 0.67, 0.8}
\definecolor{burntorange}{rgb}{0.8, 0.33, 0.0}

\lstdefinestyle{ChplStyle}{
    backgroundcolor=\color{light-gray},   
    morecomment=[l]{//},
    commentstyle=\color{burntorange},
    numberstyle=\tiny\ttfamily\bfseries,
    stringstyle=\color{mGreen},
    basicstyle=\linespread{1}\ttfamily\footnotesize,
    breakatwhitespace=false,
    xleftmargin=.1in,
    breaklines=true,                 
    captionpos=b,                    
    keepspaces=true,                 
    numbers=left,                    
    numbersep=5pt,                  
    showspaces=false,                
    showstringspaces=false,
    showtabs=false,                  
    tabsize=2,
    keywordstyle=\color{mPurple},
    deletekeywords={pragma,for,int,while,double,const,if,continue,else},
    morekeywords={pragma,omp,master,parallel,num\_threads,shared,barrier},
    keywordstyle=[2]\color{mBlue},
    keywords=[2]{var,here,type,true,false,record,int,bool,atomic,real,sync,const,class,domain,proc,double,ref,Locales}, 
  	keywordstyle=[3]\color{mRed},
    keywords=[3]{for,foreach,begin,cobegin,in,forall,coforall,do,if,else,while,reduce,on,then,continue},
}

\begin{document}

\renewcommand\thelstlisting{\arabic{lstlisting}}

\title{Compiler Optimization for Irregular Memory Access Patterns in PGAS Programs}

\author{Thomas B. Rolinger\inst{1,2}\Letter \and
Christopher D. Krieger\inst{2} \and
Alan Sussman\inst{1}}

\authorrunning{T.B. Rolinger et al.}

\institute{University of Maryland, College Park MD, USA\\
\email{\{tbrolin, als\}@cs.umd.edu}
\and Laboratory for Physical Sciences, College Park MD, USA\\
\email{krieger@lps.umd.edu}\\
\Letter\ Corresponding author: \email{tbrolin@cs.umd.edu}
}

\titlerunning{Compiler Opt. for Irregular Memory Access Patterns}

\maketitle

%
%
%
%

\begin{abstract}
Irregular memory access patterns pose performance and user productivity challenges on distributed-memory systems. 
They can lead to fine-grained remote communication and the data access patterns are often not known until runtime.
The Partitioned Global Address Space (PGAS) programming model addresses these challenges by providing users with a view of a distributed-memory system that resembles a single shared address space.
However, this view often leads programmers to write code that causes fine-grained remote communication, which can result in poor performance.
Prior work has shown that the performance of irregular applications written in Chapel, a high-level PGAS language, can be improved by manually applying optimizations.
However, applying such optimizations by hand reduces the productivity advantages provided by Chapel and the PGAS model.
We present an inspector-executor based compiler optimization for Chapel programs that automatically performs remote data replication.
While there have been similar compiler optimizations implemented for other PGAS languages, high-level features in Chapel such as implicit processor affinity lead to new challenges for compiler optimization.
We evaluate the performance of our optimization across two irregular applications.
Our results show that the total runtime can be improved by as much as 52x on a Cray XC system with a low-latency interconnect and 364x on a standard Linux cluster with an Infiniband interconnect, demonstrating that significant performance gains can be achieved without sacrificing user productivity.
\end{abstract}

%
%

\section{Introduction}
\label{sec:1_intro}
Implementing parallel software that can effectively utilize distributed-memory systems poses many challenges for programmers.
Specifically, modifying an existing serial or shared-memory parallelized application to run in a distributed setting often requires significant programmer effort to orchestrate data distribution and communication.
The Partitioned Global Address Space (PGAS) model attempts to address these challenges by providing programmers with a view of a distributed-memory system that resembles a single shared address space.
The PGAS model has been implemented in various languages and libraries, such as UPC~\cite{el2005upc}, GlobalArrays~\cite{nieplocha1996global} and Chapel~\cite{chamberlain2007parallel}.
Within PGAS languages, details regarding data distribution and communication are often abstracted from the programmer.
For example, in Chapel simply specifying an array as ``distributed'' automatically maps the data across the system and remote communication is performed implicitly.
The PGAS model therefore encourages programmers to write code in a shared-memory manner but aims to provide good performance on distributed-memory systems without requiring the program to be rewritten.

Irregular memory access patterns are commonly found in applications that perform graph analytics~\cite{lumsdaine2007challenges}, sparse linear algebra~\cite{williams2007optimization} and scientific computing operations~\cite{dongarra2016high}.
Such access patterns pose significant challenges for user productivity on distributed-memory systems because the access patterns are not known until runtime, making it difficult to orchestrate communication amongst remote processes.
However, with the PGAS model coding irregular communication patterns becomes more straightforward, as they can be implemented via one-sided communication.
Unfortunately, the performance of such codes will be significantly hindered due to the fine-grained remote communication that arises from the irregular memory accesses.
While the abstractions provided by the PGAS model can be manually bypassed to achieve better performance, such an effort would significantly degrade user productivity.

In this paper, we present the design and implementation of a compiler optimization that automatically applies the \emph{inspector-executor} technique~\cite{SALTZ91B} to parallel loops in PGAS programs written in the Chapel language.
The inspector performs memory access analysis at runtime to determine remote communication to an array of interest within a loop. 
The executor replicates the remote data and runs the original loop, but redirects remote accesses to replicated local copies to avoid repeated remote communication.
Our compiler optimization automatically identifies candidate loops and array accesses, and then performs code transformations to construct the inspector and executor routines.
As a result, the user is not required to change their original code in order to achieve significant performance gains.
The contributions of our work are as follows:
\begin{itemize}
    \item Design and implementation of an inspector-executor based compiler optimization for Chapel programs that specifically targets irregular memory accesses to distributed arrays. 
    To the best of our knowledge, this work presents the first such optimization within the Chapel compiler.
    \item Discussion on the unique features of Chapel, such as implicit processor affinity, as they relate to the compiler optimization.
    While the inspector-executor technique has been employed for a long time~\cite{SALTZ91B,IELoopParallelize,strout2003compile}, and applied to other PGAS languages~\cite{TitaniumOpt,UPCStaticDynamicCoal}, our design within Chapel requires a different approach due to Chapel's high-level features.
    \item Performance evaluation of our optimization across two irregular applications and two different distributed-memory systems.
    Our results show that the optimization can improve performance by as much as 52x on a Cray XC system with a low-latency interconnect and 364x on a standard Linux cluster with an Infiniband interconnect.
\end{itemize}

The rest of the paper is organized as follows. 
Section~\ref{sec:2_chapel} presents an overview of Chapel.
We present our compiler optimization and discuss its details in Section~\ref{sec:3_optimization}.
Section~\ref{sec:4_perfEval} presents a performance evaluation of the optimization across two irregular applications, and the Appendix contains additional performance results.
Prior work as it relates to our paper is described in Section~\ref{sec:5_related}.
Finally, Section~\ref{sec:6_concl} provides concluding remarks and discusses future work.
%
%
\section{Overview of the Chapel Language}
\label{sec:2_chapel}
Chapel is a high-level language that implements the PGAS model and is designed for productive parallel computing at scale, providing constructs for distributed arrays, remote communication and both data and task parallelism.
In this section we provide a brief overview of Chapel, focusing on the features most relevant to our work.
For a more in-depth description of Chapel, we refer readers to the work by Chamberlain et al.~\cite{chamberlain2007parallel}.

\subsection{Terminology: Tasks, Threads and Locales}
\label{sec:2-1_terms}
Chapel enables parallelism through executing multiple \emph{tasks} in parallel, where a task is a set of computations that can conceptually be executed in parallel, but may or may not do so.
Tasks are implemented by a tasking layer, which provides threads on which Chapel tasks are scheduled.
For distributed-memory programming, Chapel introduces the concept of a \emph{locale}, which is defined as a unit of machine resources on which tasks can execute.
In practice, a locale is mapped to an entire compute node in a cluster and the number of locales on which a program runs is specified when launching the program.
Chapel programs initially start with a single task executing on locale 0.
Parallel loops and other constructs then create tasks that can execute across the locales, but ultimately join back to a single task on locale 0 when they are done.
This differs from other PGAS languages, such as UPC~\cite{el2005upc}, which use a single program multiple data (SPMD) model that defines the amount of parallelism at program startup.

\subsection{Domains and Arrays}
\label{sec:2-2_domsAndArrays}
Chapel splits an array into two first-class objects in the language: a \emph{domain} and the array itself.
A domain is a representation of an index set and can be used to define the indices in an array or the indices iterated over by a loop.
Once defined, a domain can then be used to declare an array.
Modifications to a domain (e.g., adding/removing indices) propagate to all arrays defined over the domain.
Lines 1--2 in Listing~\ref{lst:domains} present a simple example of defining a domain \texttt{D} that has indices 0 through 5 and then declaring an array of integers, \texttt{data}, over that domain.
Of particular relevance to our work is an \emph{associative} domain, which is similar to a dictionary.
Lines 4--7 in Listing~\ref{lst:domains} declare an associative domain \texttt{C} that stores strings and adds the key ``foo'' via the \texttt{+=} operator.
The associative array \texttt{dict} is declared over \texttt{C} and provides a mapping of strings to reals.
Associative domains also provide parallel-safe modifications by default.

\begin{mdframed}[backgroundcolor=black!5,hidealllines=true,%
innerbottommargin=-0.85cm,innertopmargin=-0.10cm]
\noindent\begin{minipage}{\linewidth}
\begin{lstlisting}[style=ChplStyle,label={lst:domains}, caption={Chapel domains and arrays},columns=flexible]
var D = {0..5}; // rectangular domain
var data : [D] int;

var C : domain(string); // associative domain
C += "foo";
var dict : [C] real;
dict["foo"] = 2.0;
\end{lstlisting}
\end{minipage}
\end{mdframed}
\vspace{-15pt}
\begin{mdframed}[backgroundcolor=black!5,hidealllines=true,%
innerbottommargin=-0.85cm,innertopmargin=-0.1cm]
\noindent\begin{minipage}{\linewidth}
\begin{lstlisting}[style=ChplStyle,label={lst:dist}, caption={Block distributed domain/array in Chapel},columns=flexible]
var D = newBlockDom({0..15});
var arr : [D] int;

for i in 0..15 {
  arr[i] = here.id;
}
\end{lstlisting}
\end{minipage}
\end{mdframed}

The code in Listing \ref{lst:domains} only declares domains and arrays that are located on a single locale.
Our work focuses on \emph{distributed arrays}, whose data is spread across multiple locales according to some distribution policy.
Chapel provides several built-in distribution policies, such as block, cyclic and block-cyclic.
Listing~\ref{lst:dist} shows a simple example of declaring a block-distributed domain and then the corresponding distributed array.
The block distribution will partition the array into contiguous chunks and assign one chunk to each locale.
The underlying distribution implementation automatically handles the index remapping, allowing users to write \texttt{arr[i]} to access the $i^{th}$ element of the array, rather than having to specify which block the index is in.
Furthermore, Chapel performs implicit remote communication, which means that users can access the remote elements of a distributed array in the same way that they would access the local elements.
This can be seen on lines 4--6 in Listing~\ref{lst:dist}, which sets the value of \texttt{arr[i]} to be equal to the ID of the locale where the task is executing (\texttt{here.id}).
The loop will execute all iterations on locale 0, so any access to an element of \texttt{arr} that is not on locale 0 will result in communication.

\subsection{Forall Loops}
\label{sec:2-4_parallelLoops}
Chapel provides data parallelism via a \texttt{forall} loop, which allows the loop iterations to be parallelized and distributed across the system.
How the iterations are mapped to cores/locales depends on what the \texttt{forall} loop is iterating over, which is referred to as the \emph{iterand}.
Non-distributed arrays and domains have default \emph{iterators} defined, which partition the iterations into contiguous chunks and assign each chunk to a task.
The tasks then execute concurrently on a single locale.
For distributed arrays and domains, the default iterator provides both shared- and distributed-memory parallelism by executing a given iteration of the loop on the locale where that iteration's data element is mapped.
Consider the example in Listing~\ref{lst:loops}, which iterates over the domain of a distributed array \texttt{arr}.
This loop is similar to the non-parallel loop on lines 4--6 in Listing~\ref{lst:dist}, but it will execute the $i^{th}$ iteration on the locale on which \texttt{arr[i]} is stored.
Therefore, \texttt{here.id} will return the locale ID of where \texttt{arr[i]} is located.
This avoids the remote communication that would occur for the code in Listing~\ref{lst:dist}.
Therefore the iterand of a \texttt{forall} loop implicitly controls where the computation is performed.
We refer to this as controlling the \texttt{forall}'s \emph{locale affinity} and note that it is a feature of Chapel that differs from the more explicit affinity-controlling constructs of languages like UPC.
Furthermore, users can define their own custom iterators to alter the way in which a \texttt{forall} loop is parallelized.

\begin{mdframed}[backgroundcolor=black!5,hidealllines=true,%
innerbottommargin=-0.85cm,innertopmargin=-0.10cm]
\noindent\begin{minipage}{\linewidth}
\begin{lstlisting}[style=ChplStyle,label={lst:loops}, caption={\texttt{forall} loop in Chapel},columns=flexible]
var D = newBlockDom({0..15});
var arr : [D] int;
forall i in arr.domain {
    arr[i] = here.id;
}
\end{lstlisting}
\end{minipage}
\end{mdframed}

%
%

\section{Compiler Optimization}
\label{sec:3_optimization}
In this section we describe the design and implementation of the inspector-executor compiler optimization for irregular memory accesses.
We focus specifically on read-only accesses with the form \textbf{A[B[i]]} found inside \texttt{forall} loops, where \textbf{A} and \textbf{B} are arrays and \textbf{A} is a distributed array.
However, we do support more complex non-affine expressions if certain conditions are met (see Section \ref{sec:3-3_static}).
The overall goal of the optimization is to selectively replicate remotely accessed elements of \textbf{A} so that they can be accessed locally during execution of the loop.
Full replication of \textbf{A} can be prohibitively expensive in terms of both memory consumption and communication overhead, but knowing which elements are accessed remotely requires knowledge only revealed when the loop is executed.
This motivates the design of our compiler optimization that creates an inspector to determine at runtime which accesses are remote.

\begin{figure}[t]
\centering
\includegraphics[trim=0cm 6cm 0cm 0cm, scale=0.32]{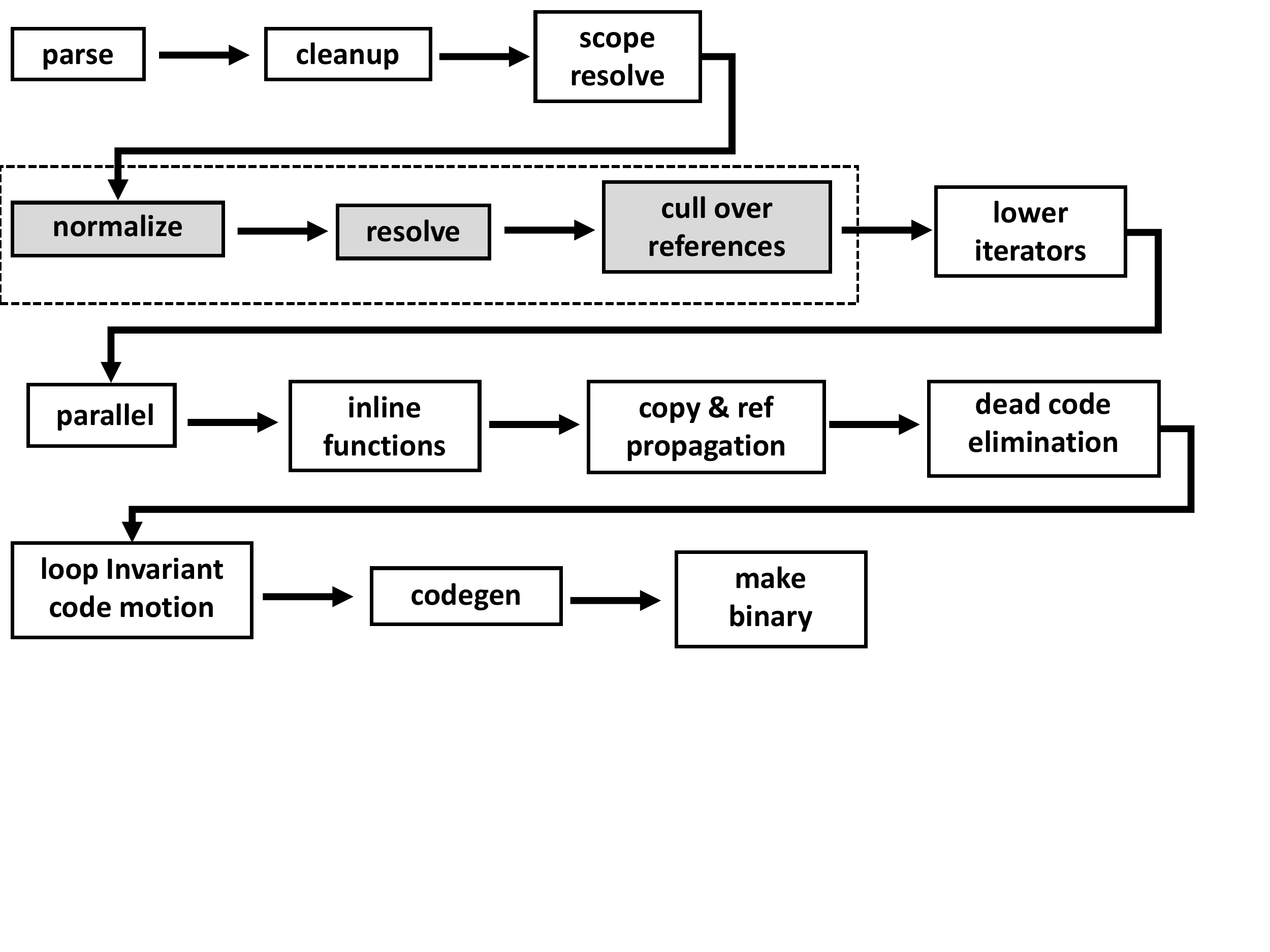}
\caption{High-level overview of the Chapel compiler. The shaded passes within the dotted box represent those where our optimization performs code transformations and static analysis.}
\label{fig:compiler}
\end{figure}

The optimization targets \texttt{forall} loops that execute multiple times, which allows for the cost of the inspector phase to be amortized over multiple executions of the optimized loop (i.e., the executor).
Such patterns are commonly found in sparse iterative solvers~\cite{dongarra2016high}, molecular dynamics simulations~\cite{strout2003compile} and some graph analytics applications~\cite{bianchini2005inside}.
Figure~\ref{fig:compiler} presents a high-level overview of Chapel's compiler passes, where the shaded passes in the dotted box correspond to the passes where the optimization performs static analysis and code transformations.
There are roughly 40 passes in Chapel's compiler to date, but most are omitted in Figure~\ref{fig:compiler} to simplify the diagram.

\subsection{High-level Overview}
\label{sec:3-1_overview}
Our optimization operates solely on Chapel's intermediate representation of the program, which we refer to as the abstract syntax tree (AST).
The optimization process begins during the normalize pass, at which point the AST has not been heavily modified, thus making the code transformation tasks easier to perform.
The optimization considers each \texttt{forall} loop that is present in the program and looks for accesses of the form \textbf{A[B[i]]}.
However, array accesses do not have their own syntax in the AST, meaning that \textbf{A[B[i]]} is represented as a call expression that is indistinguishable from a function call.
To address this issue, an irregular access candidate expression is replaced with a compiler primitive that will be acted on during the resolve pass, at which point the optimization can determine whether \textbf{A} and \textbf{B} are arrays.
If a \texttt{forall} loop has a candidate access, the optimization  clones the \texttt{forall} into the inspector and executor loops.

During the resolve pass of compilation, function calls and types are resolved, which allows the optimization to carry out much of the necessary static analysis to ensure the validity of the code transformations performed during the normalize pass.
If anything is found to invalidate the optimization then the code transformations are removed and replaced with the original \texttt{forall} loop.
Otherwise, the primitive that represents \textbf{A[B[i]]} is replaced with a library call to either ``inspect'' or ``execute'' the access, depending on whether the primitive is in the inspector or executor loop.
Determining which accesses are remote is performed at runtime via the inspector.
The static analysis performed by the optimization only determines that the access is of the form \textbf{A[B[i]]} and that \textbf{A} is a distributed array.

Finally, at the end of the cull-over references pass, the optimization performs the remaining analysis necessary to ensure the transformations are valid, which focuses on locating modifications (writes) to the relevant arrays and domains for the irregular accesses in the loop.
Such writes can change the \textbf{A[B[i]]} access pattern, which would require the inspector runtime analysis to be rerun to ensure that the executor has the correct elements replicated.
The cull-over references pass resolves the \emph{intents} of function arguments, where an argument's intent refers to whether it is passed by value or reference.
This pass statically sets the intent of the argument to different values depending on whether the procedure reads or writes to the argument.
Writes to arrays are performed via Chapel procedure calls where the array is passed in as an argument, which allows the optimization to check the intent of the array to determine whether it is written to.

\subsection{Code Transformations}
\label{sec:3-2_codeGen}
Listing~\ref{lst:codegen} presents code that is functionally equivalent to the output from our optimization's AST-level transformations when applied to the program in Listing~\ref{lst:ex}. 
We next describe the details of the code transformations and how they support selective data replication.

\begin{mdframed}[backgroundcolor=black!5,hidealllines=true,%
innerbottommargin=-0.75cm,innertopmargin=-0.10cm]
\noindent\begin{minipage}{\linewidth}
\begin{lstlisting}[style=ChplStyle,label={lst:ex}, caption={Example \texttt{forall} loop with an irregular memory access pattern},columns=flexible]
forall i in B.domain {
  C[i] = A[B[i]];
}
\end{lstlisting}
\end{minipage}
\end{mdframed}
\vspace{-10pt}
\begin{mdframed}[backgroundcolor=black!5,hidealllines=true,%
innerbottommargin=-1.1cm,innertopmargin=-0.10cm]
\noindent\begin{minipage}{\linewidth}
\begin{lstlisting}[style=ChplStyle,label={lst:codegen}, caption={Output of code transformations performed by the optimization for the code in Listing~\ref{lst:ex}},columns=flexible]
if doInspector(A, B) {
  inspectorPreamble(A);
  forall i in inspectorIterator(B.domain) {
    inspectAccess(A, B[i]);
  }
  inspectorOff(A,B);
}
executorPreamble(A);
forall i in B.domain {
  C[i] = executeAccess(A, B[i]);
}
\end{lstlisting}
\end{minipage}
\end{mdframed}

\textbf{Internal Chapel Data Structures:}
For a distributed array \textbf{A} in a \texttt{forall} loop that is being optimized, we add a Chapel record (i.e., C struct) for each locale that contains the remote communication information about \textbf{A[B[i]]} on that locale.
We refer to the complete set of these records as a \emph{communication schedule}.
A communication schedule is essentially a set of associative arrays that map \textbf{B[i]} to \textbf{A[B[i]]} when \textbf{A[B[i]]} is a remote access issued from a given locale.
If \textbf{A} is associated with multiple \texttt{forall} loops that are being optimized, the optimization will create different communication schedules for \textbf{A} that are linked to each \texttt{forall} loop.

\textbf{Inspector Loop:}
The compiler optimization starts by cloning the original \texttt{forall} loop into two copies, one that will be transformed into the inspector and one that will be transformed into the executor.
For the inspector loop (lines 3--5), the optimization replaces the original \texttt{forall} loop's iterator with a custom one (\texttt{inspectorIterator}).
This custom iterator creates one task on each locale to execute that locale's portion of the original \texttt{forall} loop serially.
While this does reduce the total amount of parallelism that the default iterator provides, it enables turning off parallel-safety for the underlying associative arrays in the communication schedule, which generally provides performance improvements.
Once the loop structure is generated, the optimization replaces the original expression \texttt{C[i] = A[B[i]]} with a call to \texttt{inspectAccess}, which is a procedure we create to perform the memory access analysis.
Note that any code in the original loop that does not pertain to the \textbf{A[B[i]]} access is removed to keep the inspector as lightweight as possible.
Furthermore, the call to \texttt{inspectAccess} does not perform the actual access to \textbf{A}, but instead queries whether \textbf{B[i]} would be a remote access to \textbf{A}, since remote accesses are expensive.
Finally, the optimization inserts a call to \texttt{inspectorPreamble} (line 2) before the inspector loop, which initializes some of the communication schedule internal structures.

\textbf{Turning the Inspector On/off:}
One of the key steps of the optimization is determining when the inspector should execute.
After the inspector runs for the first time, it does not need to be run again unless the memory access pattern changes.
The static analysis phase (in Section~\ref{sec:3-3_static}) determines when such changes occur and the optimization inserts calls to a procedure that sets flags associated with \textbf{A} and \textbf{B} to indicate that the inspector should be executed.
The \texttt{doInspector} call on line 2 checks these flags and the \texttt{inspectorOff} call on line 6 turns off the flags.

\textbf{Executor Loop:}
For the executor loop, the optimization replaces the original \textbf{A[B[i]]} access with a call to \texttt{executeAccess} (line 10), which redirects remote accesses to the replicated copies.
The optimization also inserts a call to \texttt{executorPreamble} (line 8) just before the executor loop, which initializes the replicated elements with the current values in \textbf{A}, ensuring that up-to-date values will be used in the executor.
While this does perform remote communication, the optimization only communicates a remote element once, regardless of how many times it is accessed in the loop.
As a result, the cost of the remote access is amortized over multiple local accesses.

\subsection{Static Analysis}
\label{sec:3-3_static}
There are two main goals of the static analysis: (1) to detect scenarios where the code transformations could lead to different program results compared to the original program and (2) to only apply the optimization when performance gains are likely.
Each goal imposes different requirements that must be resolved at compile-time, where violating any requirement will cause the optimization to revert to the original code.
We briefly describe the analysis that is performed to achieve these two goals.
We then discuss additional analyses that were developed to increase the generality of the optimization.

\textbf{Program Results:}
Since the optimization is applied automatically by the compiler, it must detect scenarios where the code transformations could produce program results that differ from the original code.
The primary scenario that could lead to different program results is not running the inspector when necessary, resulting in communication schedules that are out of date.
To determine when the inspector needs to execute, the optimization must be able to statically reason about the locale affinity of the \texttt{forall} loop and the source/destination locales of the remote accesses for \textbf{A[B[i]]}.
The following criteria summarize the static checks performed for these purposes:
\begin{enumerate}
    \item The \texttt{forall} loop must iterate over a distributed array or distributed domain.
    \item The \texttt{forall} loop cannot be nested inside of another \texttt{forall} or any other statement that could create multiple parallel tasks.
    \item The index \textbf{i} in \textbf{A[B[i]]} must be the loop index variable for the loop that contains \textbf{A[B[i]]}, and the loop must iterate over a domain or array.
    \item Neither \textbf{A} nor \textbf{B}, nor their domains, can be modified within the \texttt{forall} loop.
\end{enumerate}
Check (1) ensures that the optimization can reason about the \texttt{forall} loop's locale affinity (i.e., which locales the iterations will execute on).
Check (2) ensures that multiple tasks will not execute the entire \texttt{forall} loop at the same time, which would result in potential concurrent updates to the communication schedules. 
Check (3) ensures that the optimization can statically reason about
the index \textbf{i} into \textbf{B}, specifically when the values of \textbf{i} would change. 
Such changes would happen if the array/domain over which the loop iterates is modified.
Check (4) ensures that the values in \textbf{B} that are analyzed by the inspector will be the same as the values used within the executor (and the original loop).
When applied to \textbf{A}, this check avoids the complexities of writing to replicated elements of \textbf{A} and having to propagate those values back to the original elements.
If these checks are met, then the optimization will determine all modifications to \textbf{A}, \textbf{B}, their domains and the various other arrays/domains involved in the loop.
If these objects are modified, it indicates that \textbf{A[B[i]]} could exhibit an access pattern that differs from when the inspector was last performed, whether it be different indices used to access \textbf{A} or the accesses themselves being issued from different locales.
When modifications are found, the optimization will set the corresponding flags to rerun the inspector to update the communication schedule.

\textbf{Program Performance:}
The goal of the optimization is to improve the runtime performance of an input program.
Therefore the static analysis attempts to determine whether the optimization is likely to provide performance gains.
This analysis can be summarized as determining that the \texttt{forall} loop will execute multiple times without requiring the inspector to be executed each time, which would incur significant overhead.
The following criteria summarize the static checks performed for these purposes:
\begin{enumerate}[(a)]
    \item The \texttt{forall} loop must be nested in an outer serial loop (i.e., \texttt{for}, \texttt{while}, etc.).
    \item Neither \textbf{B} nor its domain can be modified within the outer loop that the \texttt{forall} is nested in.
    \item \textbf{A}'s domain cannot be modified within the outer loop that the \texttt{forall} is nested in.
\end{enumerate}
Check (a) ensures that the \texttt{forall} loop is likely to be executed multiple times, though it is not guaranteed to do so.
Checks (b) and (c) ensure that the inspector will not be executed each time the \texttt{forall} loop runs.
Recall that part of the code transformation phase is to ``turn on'' the inspector after modifications to \textbf{B}, its domain or \textbf{A}'s domain.
Such modifications have the potential to alter the memory access pattern of \textbf{A[B[i]]}, and therefore require the inspector to be executed again.
The array \textbf{A} is allowed to be modified within the outer loop, as any changes to its values will be propagated to the replicated copies via the \texttt{executorPreamble}.
However, modifying \textbf{A}'s domain could alter the access pattern \textbf{A[B[i]]} by adding/removing elements in \textbf{A}.

\textbf{Non-affine Expression Analysis:}
Thus far we have focused our discussion on accesses of the form \textbf{A[B[i]]}.
However, the optimization can support more complex non-affine expressions, such as \textbf{A[B[i*j]\%k+1]}, if certain conditions are met.
Specifically, the expressions must be binary operations between immediates or variables yielded by loops that iterate over arrays/domains.
This requirement is needed so that the optimization can statically reason about when/how the accesses to \textbf{A} and \textbf{B} could change.
This is accomplished by locating modifications to the arrays/domains that yield the variables used in the expressions.
Note that outside of our optimization, Chapel's current compiler does not perform affine/non-affine expression analysis.

\textbf{Interprocedural and Alias Analyses:}
The optimization performs interprocedural and alias analyses to support the static checks described previously.
The interprocedural analysis computes the call graph starting from the function that contains the \texttt{forall} and checks for any invalid call paths.
We deem a call path invalid if there is a lack of an outer serial loop or if there is an enclosing statement that creates multiple parallel tasks.
When an invalid call path is detected the compiler inserts flags that are set at runtime to ``turn off'' the optimization temporarily along the invalid path.
Also, the interprocedural analysis detects modifications to the arrays/domains of interest across arbitrarily nested function calls.
Alias analysis is necessary because Chapel allows users to create references to arrays/domains, which operate similarly to pointers in C.
To address this issue, we developed static checks to detect such references and determine the original array/domain.
This analysis works for arbitrarily long alias chains (e.g., \texttt{var arr = ...; ref a1 = arr; ref a2 = a1}), where the optimization will detect modifications to any of the references along the chain.
%
%
\section{Performance Evaluation}
\label{sec:4_perfEval}
To demonstrate the performance benefits of the compiler optimization described in Section~\ref{sec:3_optimization}, we performed an evaluation of two irregular applications running on two different distributed-memory systems.
The applications we evaluate are implemented in a high-level manner, consistent with Chapel's design philosophy, which is to separate data distribution/communication details from the algorithm design.
As a result, our exemplar applications closely match standard shared-memory implementations and are representative of direct use of the PGAS model.
Our goal is to show that the performance of these Chapel programs suffer from implicit fine-grained remote communication, but can be significantly improved via automatic optimization without requiring the user to modify the program.
Therefore, users can take advantage of the productivity benefits that Chapel provides while also achieving good performance.

\subsection{Experimental Setup}
\label{sec:4-1_setup}
For our evaluation, we run experiments on a Cray XC cluster and an Infiniband-based cluster.
For the Cray XC, we utilize up to 64 nodes connected over an Aries interconnect, where each node has two 22-core Intel Xeon Broadwell CPUs and 128~GB of DDR4 memory.
For the Infiniband cluster, we utilize up to 32 nodes connected over an FDR Infiniband interconnect, where each node has two 10-core Intel Xeon Haswell CPUs and 512~GB of DDR4 memory.
On the Cray XC, Chapel is built using the ugni communication layer and the \texttt{aries} communication substrate.
On the Infiniband system, Chapel is built using the GASNet communication layer and the \texttt{ibv} communication substrate.
All applications on both systems are compiled using the \texttt{---fast} flag.
For each experiment, we execute the given application multiple times and measure the total runtime, including the inspector overhead.
The results represent the average of these trials.
We observed that the runtime variation between trials did not exceed 4\%.

\subsection{Application: NAS-CG}
\label{sec:4-2_cg}
\begin{mdframed}[backgroundcolor=black!5,hidealllines=true,%
innerbottommargin=-0.75cm,innertopmargin=-0.10cm]
\noindent\begin{minipage}{\linewidth}
\begin{lstlisting}[style=ChplStyle,label={lst:cg}, caption={\texttt{forall} loop for NAS-CG},columns=flexible]
forall row in Rows {
  var accum : real = 0;
  for k in row.offsets {
    accum += values[k] * x[col_idx[k]];
  }
  b[row.id] = accum;
}
\end{lstlisting}
\end{minipage}
\end{mdframed}
The conjugate gradient (CG) method solves the equation $Ax=b$ for $x$, where $A$ is a symmetric positive-definite matrix and is typically large and sparse, and $x$ and $b$ are vectors.
Unstructured optimization problems and partial differential equations can be solved using iterative CG methods.
For the evaluation, we use the NAS-CG benchmark specification and datasets~\cite{bailey1995parallel}.
Table~\ref{tab:cg} describes the problem sizes evaluated, where each problem size corresponds to the size of the $A$ matrix.
Each iteration of NAS-CG performs a total of 26 sparse matrix-vector multiplies (SpMVs), which is the kernel of interest for the optimization and is shown in Listing~\ref{lst:cg}.
The implementation uses a standard Compressed Sparse Row (CSR) format to represent $A$, where \texttt{Rows} is a block distributed array of records that contains the offsets into the distributed array(s) containing the non-zero data values.
This approach closely resembles the Fortan+OpenMP implementation of NAS-CG~\cite{bailey1995parallel}.
The irregular access of interest is on line 4, \texttt{x[col\_idx[k]]}.
For the optimization, the inspector only needs to be executed once since the memory access pattern remains the same across all SpMV operations.
In regards to memory storage overhead due to replication, we observed an average increase in memory usage of 6\%.

%
%
\vspace{-10mm}
\begin{table}
\renewcommand{\arraystretch}{1.0}
\caption{Datasets for NAS-CG}
\label{tab:cg}
\centering
\begin{tabular}{|c|c|c|c|c|}
\hline
\textbf{Name} & \textbf{Rows} & \textbf{Non-zeros} & \textbf{Density (\%)} & \textbf{\# of SpMVs} \\
\hline
C           & 150k & 39M  & 0.17  & 1950  \\
D           & 150k & 73M  & 0.32  & 2600 \\ 
E           & 9M   & 6.6B & 0.008 & 2600 \\
F           & 54M  & 55B  & 0.002 & 2600\\
\hline
\end{tabular}
\caption{Runtime speed-ups achieved by the optimization on the NAS-CG problem sizes from Table~\ref{tab:cg} relative to the unoptimized code in Listing \ref{lst:cg}. Missing values indicate that the problem size required too much memory to execute and ``NA'' values indicate that the system did not support the specified number of nodes/locales.}
\label{tab:nascg_speedups}
\centering
\begin{tabular}{|c|cccc|cccc|}
\hline
& \multicolumn{4}{c|}{\textbf{Cray XC}} & \multicolumn{4}{c|}{\textbf{Infiniband}} \\
\textbf{Locales (Nodes)} & C   & D   & E    & F    & C    & D    & E   & F   \\ \hline
2                & 3.2 & 2.8 & ---  & ---  & 8.9  & 6    & 357 & --- \\
4                & 3.6 & 3.4 & 17.5 & ---  & 15.8 & 10.4 & 345 & --- \\
8                & 5.7 & 6.2 & 36.7 & ---  & 115  & 127  & 364 & --- \\
16               & 8.6 & 11  & 22.5 & ---  & 238  & 330  & 258 & 270 \\
32               & 6.4 & 8.4 & 34   & 52.3 & 160  & 240  & 195 & 165 \\
64               & 4.1 & 4.9 & 16.7 & 25.4 & NA   & NA   & NA  & NA  \\
\hline
\textbf{geomean} & 5   & 5.5 & 24.1 & 36.4 & 57.3 & 57.5 & 296 & 211 \\
\hline
\end{tabular}
\end{table}
\vspace{-5mm}

Table~\ref{tab:nascg_speedups} presents the NAS-CG runtime speed-ups achieved by the optimization on each system relative to the unoptimized code shown in Listing~\ref{lst:cg}.
We observe large speed-ups on both systems, but most notably on the Infiniband system because fine-grained remote communication exhibits higher latency compared to the Aries interconnect on the Cray XC.
On both systems, such speed-ups are obtained because of a large amount of remote data reuse in the SpMV kernel, which is due to the sparsity pattern of the matrices generated by the benchmark.
The optimization incurs the cost of reading the remote element once, but can then access the element locally throughout the rest of the \texttt{forall} loop.
Without the optimization, each access to the remote element likely pays the full latency cost of a remote access since the access pattern is sparse and irregular.
Additionally, the SpMV kernel is executed many times during the NAS-CG benchmark, which allows for the inspector overhead to be amortized.
We observe that the percentage of runtime devoted to the inspector is 3\% on average across all locale counts for the Cray XC and 2\% for the Infiniband system.

Beyond relative speed-ups, the optimization significantly improves the overall runtime, as can be see in Tables~\ref{tab:nascg_base_times} and~\ref{tab:nascg_opt_times} in the Appendix.
We observe very poor runtime performance and scalability for the unoptimized code, which is due to the implicit fine-grained remote communication required in the straightforward implementation shown in Listing~\ref{lst:cg}.
On the other hand, the optimized version of the code generally gets faster with more locales, since the optimization can take advantage of remote data reuse.
Overall, these results demonstrate the usefulness of our optimization in automatically providing faster runtimes without sacrificing user productivity.

\subsection{Application: PageRank}
\label{sec:4-3_pr}
\begin{mdframed}[backgroundcolor=black!5,hidealllines=true,%
innerbottommargin=-0.75cm,innertopmargin=-0.10cm]
\noindent\begin{minipage}{\linewidth}
\begin{lstlisting}[style=ChplStyle,label={lst:pr}, caption={\texttt{forall} loop for PageRank},columns=flexible]
forall v in Graph {
  var val = 0.0;
  for i in v.offsets {
    ref t = Graph[neighbors[i]];
    val += t.pr_read / t.out_degree;
  }
  v.pr_write = (val * d) + ((1.0-d)/num_vertices) + sink_val;
}
\end{lstlisting}
\end{minipage}
\end{mdframed}

PageRank~\cite{bianchini2005inside} is an iterative graph algorithm that provides an importance measurement for each vertex in a graph.
Listing~\ref{lst:pr} presents the PageRank kernel and is the target of the optimization, where the irregular memory access of interest is on line 4.
Similar to NAS-CG, we use a CSR data structure to represent the graph.
The distributed array \texttt{Graph} stores records that correspond to vertices, where each vertex has two importance measurements: \texttt{pr\_write} and \texttt{pr\_read}.
This allows for a straightforward parallel implementation of the kernel by treating one value as read-only during an iteration, and closely matches the GAP Benchmark Suite implementation~\cite{GAPBS}.
Unlike NAS-CG, PageRank adds the complication of storing records in the array of interest rather than base type data (i.e., \texttt{int}, \texttt{real}, etc.).
The optimization recognizes this feature and will only replicate the fields that are accessed in the \texttt{forall} loop, namely, \texttt{pr\_read} and \texttt{out\_degree}.

We evaluate PageRank on two real web graphs obtained from the SuiteSparse Matrix Collection~\cite{MatrixMarket}, which are described in Table~\ref{tab:pr}.
The right-most column denotes the number of iterations that are required to converge with a tolerance value of $1e{\text -}7$ and a damping factor of 0.85 (\texttt{d} on line 7 in Listing \ref{lst:pr}).
Our choice for these values matches what is used in Neo4j~\cite{neo4j}, an open source graph database.
Each PageRank iteration performs one execution of the entire \texttt{forall} in Listing~\ref{lst:pr} and the inspector is only executed once since the graph does not change throughout the execution.
The memory storage overhead of the optimization for PageRank is 40--80\%, which is much larger than what we observed for NAS-CG.
For NAS-CG, the array of interest that is replicated constitutes a small portion of the total memory required.
But for PageRank, the array of interest is much larger by comparison, resulting in a larger relative increase in memory storage.
However, the memory storage overhead incurred by the optimization is significantly less than what full replication would incur.

%
%
\vspace{-10mm}
\begin{table}
\renewcommand{\arraystretch}{1.0}
\caption{Datasets for PageRank}
\label{tab:pr}
\centering
\begin{tabular}{|c|c|c|c|c|}
\hline
\textbf{Name} & \textbf{Vertices} & \textbf{Edges} & \textbf{Density (\%)} & \textbf{Iterations}\\
\hline
webbase-2001    & 118M & 992M & \num{7.1e-6} & 33 \\
sk-2005         & 51M  & 1.9B & \num{7.5e-5} & 40 \\
\hline
\end{tabular}
\end{table}

\begin{table}[h]
\renewcommand{\arraystretch}{1}
\caption{Runtime speed-ups achieved by the optimization on PageRank for the graphs from Table~\ref{tab:pr} relative to the unoptimized implementation. ``NA'' values indicate that the system did not support the specified number of nodes/locales.}
\label{tab:pr_speedups}
\centering
\begin{tabular}{|c|cc|cc|}
\hline
& \multicolumn{2}{c|}{\textbf{Cray XC}} & \multicolumn{2}{c|}{\textbf{Infiniband}} \\
\textbf{Locales (Nodes)} & webbase-2001  & sk-2005 & webbase-2001 & sk-2005 \\ \hline
2                & 0.88          & 1.2     & 5.2          & 2   \\
4                & 0.98          & 1.6     & 8.6          & 7.1 \\
8                & 0.97          & 1.3     & 12           & 6   \\
16               & 0.94          & 1.7     & 9.6          & 5.4 \\
32               & 1.3           & 1.4     & 4.5          & 4.2 \\
64               & 1.2           & 2.1     & NA           & NA  \\
\hline
\textbf{geomean} & 1.04          & 1.5     & 7.3          & 4.5 \\
\hline
\end{tabular}
\end{table}
\vspace{-10pt}

Table~\ref{tab:pr_speedups} presents the runtime speed-ups for the optimization on both graphs from Table~\ref{tab:pr}, and Tables~\ref{tab:pr_base_times} and~\ref{tab:pr_opt_times} in the Appendix presents the runtimes without and with the optimization.
As we observed for NAS-CG, the speed-ups on the Infiniband system are larger than those on the Cray XC due to the Aries interconnect.
However, we observe significantly smaller speed-ups overall when compared to NAS-CG.
This is largely because the PageRank kernel is executed fewer times than the NAS-CG kernel, and the graphs exhibit significantly less data reuse when compared to NAS-CG.
As a result, there is a larger inspector overhead and smaller performance gains from the executor.
For these reasons, speed-ups are not achieved on the webbase-2001 graph on the Cray XC until 32 locales, which is when the remote data reuse reaches its peak.
Furthermore, due to the highly irregular nature of the graphs (whose degree distributions follow a power law), the runtime performance fluctuates as the number of locales increase due to the partitioning of the graph across the system.
This can change where the elements are located, which may lead to a once heavily accessed remote element now being local.
This is more significant for the unoptimized code, as the optimization would have only incurred the cost of the remote access once due to replication.
Nevertheless, the scalability of the optimization generally tracks the scalability of the code without the optimization applied.
%
%
\section{Related Work}
\label{sec:5_related}
Techniques to perform runtime optimizations for irregular memory accesses have been worked on for many years, and of particular relevance to our work is the \emph{inspector-executor} technique~\cite{SALTZ91B,KenImprovingCachePerformance,strout2003compile}.
Das et al.~\cite{das1994communication} presented an inspector-executor optimization similar to ours, but that work predates the PGAS model, so is not directly applicable due to fundamental differences in programming model design and implementation.
For PGAS languages, Su and Yelick~\cite{TitaniumOpt} developed an inspector-executor optimization that is similar to ours but for the language Titanium. 
However, Titanium differs from Chapel in its execution model as well as its language constructs for parallel loops, since it is based on Java.
This leads to an overall different approach to static analysis.
Alvanos et al.~\cite{UPCStaticDynamicCoal,ALVANOS20162} described an inspector-executor framework for the PGAS language UPC~\cite{el2005upc}, which also differs from Chapel.
UPC uses a SPMD model and requires explicit constructs to control the processor affinity of parallel loops.
Like Titanium, these differences lead to a significantly different approach to the static analysis used for the optimization.
However, we plan to explore some of their techniques in future work.
For compiler optimizations related specifically to Chapel, Kayraklioglu et al.~\cite{enginLCPC2021} presented an optimization to aggregate remote accesses to distributed arrays.
However, their optimization does not specifically target irregular memory accesses, which leads to significantly different approaches to static analysis and code transformation.
Furthermore, the applications we evaluate in Section~\ref{sec:4_perfEval} are not candidates for their aggregation optimization.
%
%
\section{Conclusions and Future Work}
\label{sec:6_concl}
While the PGAS model provides user productivity advantages for writing distributed irregular applications, the resulting code often has poor runtime performance due to fine-grained remote communication.
In this work we have presented a compiler optimization for Chapel programs that specifically targets irregular memory access patterns within parallel loops and automatically applies code transformations to replicate remotely accessed data.
We demonstrated that the optimization provides runtime speed-ups as large as 52x on a Cray XC system and 364x on an Infiniband system.
To this end, we have shown that significant performance gains can be achieved without sacrificing user productivity.

For future work, we plan to improve upon our compiler optimization framework and address some of the limitations, such as optimizing multiple irregular accesses in the same loop.
We also plan to design additional compiler optimizations to serve as alternatives to selective data replication when that cannot be applied, as it currently is only applicable for read-only data.
Future optimizations will specifically target writes.

\section*{Acknowledgements}
We would like to thank Brad Chamberlain, Engin Kayraklioglu, Vass Litvinov, Elliot Ronaghan and Michelle Strout from the Chapel team for their guidance on working with the Chapel compiler, as well as providing access to the Cray XC system that was used in our performance evaluation.
%
%
\section*{Appendix}
\label{sec:7_appendix}
In this appendix we provide additional performance results for the experiments performed in Section~\ref{sec:4_perfEval}.
Table~\ref{tab:nascg_base_times} presents execution runtimes for the unoptimized implementation of NAS-CG and Table \ref{tab:nascg_opt_times} presents the runtimes for the automatically optimized implementation of NAS-CG.
Due to the amount of time required by the unoptimized code for problem sizes E and F, we project their total runtime based on their average iteration runtime.
Each iteration of NAS-CG performs the same computation and communication, and we note that the variation between iterations is no more than 2\% on problem sizes C and D.
Tables~\ref{tab:pr_base_times} and~\ref{tab:pr_opt_times} present similar data for the PageRank application.

%
%
\begin{table}
\caption{Execution runtimes (in minutes) for the unoptimized implementation of NAS-CG when executed on the problem sizes from Table~\ref{tab:cg}. Missing values indicate that the problem size required too much memory to execute and ``NA'' values indicate that the system did not support the specified number of nodes/locales. Runtimes for problem sizes E and F are projected from single iteration runtimes.}
\label{tab:nascg_base_times}
\centering
\begin{tabular}{|c|cccc|cccc|}
\hline
& \multicolumn{4}{c|}{\textbf{Cray XC}} & \multicolumn{4}{c|}{\textbf{Infiniband}} \\
\textbf{Locales (Nodes)} & C   & D   & E    & F    & C  & D   & E                   & F   \\ \hline
2                & 1.8 & 3.9 & ---  & ---  & 9  & 15  & $2.3\mathrm{e}{+5}$ & --- \\
4                & 1.7 & 3.4 & 2882 & ---  & 12 & 17  & $1.7\mathrm{e}{+5}$ & --- \\
8                & 2.2 & 4.8 & 3069 & ---  & 58 & 131 & $1\mathrm{e}{+5}$   & --- \\
16               & 2.7 & 6.1 & 1005 & ---  & 96 & 242 & $5.7\mathrm{e}{+4}$ & $4.7\mathrm{e}{+5}$ \\
32               & 1.8 & 4   & 845  & 9927 & 64 & 156 & $3.1\mathrm{e}{+4}$ & $2.4\mathrm{e}{+5}$ \\
64               & 1.2 & 2.6 & 285  & 3501 & NA & NA  & NA                  & NA  \\
\hline
\end{tabular}
\caption{Execution runtimes (in minutes) for the automatically optimized implementation of NAS-CG when executed on the problem sizes from Table~\ref{tab:cg}. Missing values indicate that the problem size required too much memory to execute and ``NA'' values indicate that the system did not support the specified number of nodes/locales. Runtimes for problem sizes E and F are projected from single iteration runtimes.}
\label{tab:nascg_opt_times}
\centering
\begin{tabular}{|c|cccc|cccc|}
\hline
& \multicolumn{4}{c|}{\textbf{Cray XC}} & \multicolumn{4}{c|}{\textbf{Infiniband}} \\
\textbf{Locales (Nodes)} & C   & D   & E   & F   & C   & D   & E   & F    \\ \hline
2                & 0.6 & 1.4 & --- & --- & 1.1 & 2.6 & 655 & ---  \\
4                & 0.5 & 1   & 165 & --- & 0.7 & 1.7 & 489 & ---  \\
8                & 0.4 & 0.8 & 84  & --- & 0.5 & 1   & 279 & ---  \\
16               & 0.3 & 0.6 & 45  & --- & 0.4 & 0.7 & 221 & 1732 \\
32               & 0.3 & 0.5 & 25  & 190 & 0.4 & 0.6 & 156 & 1437 \\
64               & 0.3 & 0.5 & 17  & 138 & NA  & NA  & NA  & NA   \\
\hline
\end{tabular}
\end{table}

\begin{table}
\caption{Execution runtimes (in minutes) for the unoptimized implementation of PageRank when executed on the graphs from Table~\ref{tab:pr}. ``NA'' values indicate that the system did not support the specified number of nodes/locales.}
\label{tab:pr_base_times}
\centering
\begin{tabular}{|c|cc|cc|}
\hline
& \multicolumn{2}{c|}{\textbf{Cray XC}} & \multicolumn{2}{c|}{\textbf{Infiniband}} \\
\textbf{Locales (Nodes)} & webbase-2001 & sk-2005 & webbase-2001 & sk-2005 \\ \hline
2                & 1.1           & 4.1    & 14.6         & 14   \\
4                & 0.9           & 3.8    & 14.6         & 28 \\
8                & 0.7           & 4.4    & 14.9         & 23   \\
16               & 0.6           & 7.3    & 9.7          & 29 \\
32               & 0.6           & 5.7    & 6.9          & 26 \\
64               & 0.7           & 10     & NA           & NA  \\
\hline
\end{tabular}

\caption{Execution runtimes (in minutes) for the automatically optimized implementation of PageRank when executed on the graphs from Table~\ref{tab:pr}. ``NA'' values indicate that the system did not support the specified number of nodes/locales.}
\label{tab:pr_opt_times}
\centering
\begin{tabular}{|c|cc|cc|}
\hline
& \multicolumn{2}{c|}{\textbf{Cray XC}} & \multicolumn{2}{c|}{\textbf{Infiniband}} \\
\textbf{Locales (Nodes)} & webbase-2001 & sk-2005 & webbase-2001 & sk-2005 \\ \hline
2                & 1.2          & 3.3     & 2.8          & 6.9   \\
4                & 1            & 2.4     & 1.7          & 3.9 \\
8                & 0.8          & 3.4     & 1.2          & 3.9   \\
16               & 0.7          & 4.2     & 1.1          & 5.5 \\
32               & 0.5          & 4.2     & 1.5          & 6.2 \\
64               & 0.6          & 4.7     & NA           & NA  \\
\hline
\end{tabular}
\end{table}

\newpage
\clearpage
\bibliographystyle{splncs04}
\bibliography{refs}

\begin{thebibliography}{10}
\providecommand{\url}[1]{\texttt{#1}}
\providecommand{\urlprefix}{URL }
\providecommand{\doi}[1]{https://doi.org/#1}

\bibitem{UPCStaticDynamicCoal}
Alvanos, M., Farreras, M., Tiotto, E., Amaral, J.N., Martorell, X.: Improving
  communication in {PGAS} environments: static and dynamic coalescing in {UPC}.
  In: Proceedings of the 27th International ACM Conference on Supercomputing
  (ICS'13). p. 129–138. Association for Computing Machinery (2013).
  \doi{10.1145/2464996.2465006}, \url{https://doi.org/10.1145/2464996.2465006}

\bibitem{ALVANOS20162}
Alvanos, M., Tiotto, E., Amaral, J.N., Farreras, M., Martorell, X.: Using
  shared-data localization to reduce the cost of inspector-execution in
  {Unified-Parallel-C} programs. Parallel Computing  \textbf{54},  2--14
  (2016). \doi{https://doi.org/10.1016/j.parco.2016.03.002}

\bibitem{bailey1995parallel}
Bailey, D., Harris, T., Saphir, W., Van Der~Wijngaart, R., Woo, A., Yarrow, M.:
  The {NAS} parallel benchmarks 2.0. Tech. rep., Technical Report NAS-95-020,
  NASA Ames Research Center (1995)

\bibitem{GAPBS}
Beamer, S., Asanović, K., Patterson, D.: The {GAP} benchmark suite (2015).
  \doi{10.48550/ARXIV.1508.03619}

\bibitem{bianchini2005inside}
Bianchini, M., Gori, M., Scarselli, F.: Inside {PageRank}. ACM Transactions on
  Internet Technology  \textbf{5}(1),  92--128 (2005)

\bibitem{chamberlain2007parallel}
Chamberlain, B.L., Callahan, D., Zima, H.P.: Parallel programmability and the
  {Chapel} language. The International Journal of High Performance Computing
  Applications  \textbf{21}(3),  291--312 (2007)

\bibitem{das1994communication}
Das, R., Uysal, M., Saltz, J., Hwang, Y.S.: Communication optimizations for
  irregular scientific computations on distributed memory architectures.
  Journal of Parallel and Distributed Computing  \textbf{22}(3),  462--478
  (1994)

\bibitem{MatrixMarket}
Davis, T.A., Hu, Y.: The {University of Florida Sparse Matrix Collection}. ACM
  Trans. Math. Softw.  \textbf{38}(1),  1:1--1:25 (Dec 2011).
  \doi{10.1145/2049662.2049663}

\bibitem{KenImprovingCachePerformance}
Ding, C., Kennedy, K.: Improving cache performance in dynamic applications
  through data and computation reorganization at run time. In: Proceedings of
  the ACM SIGPLAN 1999 Conference on Programming Language Design and
  Implementation. pp. 229--241. ACM (1999). \doi{10.1145/301618.301670}

\bibitem{dongarra2016high}
Dongarra, J., Heroux, M.A., Luszczek, P.: High-performance conjugate-gradient
  benchmark: A new metric for ranking high-performance computing systems. The
  International Journal of High Performance Computing Applications
  \textbf{30}(1),  3--10 (2016)

\bibitem{el2005upc}
El-Ghazawi, T., Carlson, W., Sterling, T., Yelick, K.: UPC: Distributed Shared
  Memory Programming, Wiley Series on Parallel and Distributed Computing,
  vol.~40. John Wiley \& Sons (2005)

\bibitem{enginLCPC2021}
Kayraklioglu, E., Ronaghan, E., Ferguson, M.P., Chamberlain, B.L.:
  Locality-based optimizations in the {Chapel} compiler. In: Li, X.,
  Chandrasekaran, S. (eds.) Languages and Compilers for Parallel Computing. pp.
  3--17. Springer (2021)

\bibitem{lumsdaine2007challenges}
Lumsdaine, A., Gregor, D., Hendrickson, B., Berry, J.: Challenges in parallel
  graph processing. Parallel Processing Letters  \textbf{17}(01),  5--20 (2007)

\bibitem{nieplocha1996global}
Nieplocha, J., Harrison, R.J., Littlefield, R.J.: {Global Arrays}: A nonuniform
  memory access programming model for high-performance computers. The Journal
  of Supercomputing  \textbf{10}(2),  169--189 (1996)

\bibitem{IELoopParallelize}
Rauchwerger, L., Padua, D.: The {LRPD} test: speculative run-time
  parallelization of loops with privatization and reduction parallelization.
  IEEE Transactions on Parallel and Distributed Systems  \textbf{10}(2),
  160--180 (1999). \doi{10.1109/71.752782}

\bibitem{SALTZ91B}
Saltz, J.H., Mirchandaney, R., Crowley, K.: Run-time parallelization and
  scheduling of loops. IEEE Transactions on Computers  \textbf{40}(5),
  603--612 (1991)

\bibitem{strout2003compile}
Strout, M.M., Carter, L., Ferrante, J.: Compile-time composition of run-time
  data and iteration reorderings. In: Proceedings of the ACM SIGPLAN 2003
  Conference on Programming language Design and Implementation. pp. 91--102.
  ACM (2003)

\bibitem{TitaniumOpt}
Su, J., Yelick, K.: Automatic support for irregular computations in a
  high-level language. In: IEEE International Parallel and Distributed
  Processing Symposium. vol.~2, pp. 53b--53b. IEEE (2005)

\bibitem{neo4j}
Webber, J.: A programmatic introduction to {Neo4j}. In: Proceedings of the 3rd
  Annual Conference on Systems, Programming, and Applications: Software for
  Humanity. p. 217–218. ACM (2012). \doi{10.1145/2384716.2384777}

\bibitem{williams2007optimization}
Williams, S., Oliker, L., Vuduc, R., Shalf, J., Yelick, K., Demmel, J.:
  Optimization of sparse matrix-vector multiplication on emerging multicore
  platforms. In: SC'07: Proceedings of the 2007 ACM/IEEE Conference on
  Supercomputing. pp. 1--12. IEEE (2007)

\end{thebibliography}

\end{document}